# Photonic band gap in isotropic hyperuniform disordered solids with low dielectric contrast

Weining Man,[1*] Marian Florescu,[2] Kazue Matsuyama,[1] Polin Yadak,[1] Geev Nahal,[1] Seyed Hashemizad,[1] Eric Williamson,[1] Paul Steinhardt,[3,4] Salvatore Torquato,[3,4,5] and Paul Chaikin[6]

[1]*Department of Physics and Astronomy, San Francisco State University, San Francisco, CA 94132, USA*
[2]*Advanced Technology Institute, Physics Department, University of Surrey, Guildford, Surrey GU2 7XH, UK*
[3]*Department of Physics, Princeton University, Princeton, NJ 08544, USA*
[4]*Princeton Center for Theoretical Science, Princeton University, Princeton, NJ 08544, USA*
[5]*Department of Chemistry , Princeton University, Princeton, NJ 08544, USA*
[6]*Department of Physics, New York University, New York, NY 20012, USA*
[*]*weining@sfsu.edu*

**Abstract:** We report the first experimental demonstration of a TE-polarization photonic band gap (PBG) in a 2D isotropic hyperuniform disordered solid (HUDS) made of dielectric media with a index contrast of 1.6:1, very low for PBG formation. The solid is composed of a connected network of dielectric walls enclosing air-filled cells. Direct comparison with photonic crystals and quasicrystals permitted us to investigate band-gap properties as a function of increasing rotational isotropy. We present results from numerical simulations proving that the PBG observed experimentally for HUDS at low index contrast has zero density of states. The PBG is associated with the energy difference between complementary resonant modes above and below the gap, with the field predominantly concentrated in the air or in the dielectric. The intrinsic isotropy of HUDS may offer unprecedented flexibilities and freedom in applications (i. e. defect architecture design) not limited by crystalline symmetries.

**1. Introduction.**

PGB materials, first introduced in 1987 [1, 2], have drawn great interest both for their scientific properties and for their technological applications [3-6]. Until recently, most PBG materials have been periodic (crystalline) structures exhibiting Bragg scattering peaks. In a one-dimensional (1D) periodic structure, a PBG exists for arbitrarily low dielectric contrast. However, in two- or three-dimensional periodic structures, only a limited number of discrete rotational symmetries are possible according to the fundamental theorems of crystallography. The crystal's anisotropy results in different stop bands (blocking frequencies) in different directions and makes it difficult to form a PBG in higher-dimensional structures in the absence of a large dielectric contrast. Quasicrystalline structures can have a higher degree of symmetry than crystalline structures, helping to form a band gap (covering all directions), as suggested in [7-9].

Disordered photonic materials have attracted increasing attention in the recent years due to their broadband and wide-angle properties [10]. Recent numerical and experimental work [11-14] has shown that periodicity or quasiperiodicity and the associated Bragg-diffraction peaks are not a prerequisite for PBG formation. Simulations in [13] predicted that sizeable complete (for all polarizations) 2D PBGs can exist in a new class of "designer" hyperuniform disordered structures with *high* dielectric contrast ($\varepsilon$=11.5), a connected-wall network decorated with dielectric cylinders. The presence of a complete PBG in a hyperuniform disordered structure decorated with alumina ($\varepsilon$=8.76) was recently demonstrated by Man *et. al.* [14].

Most single-polarization PBG studies in photonic crystals with disorder have been conducted only for TM polarization using lattices decorated with individual dielectric cylinders. Disorder was conventionally found to wipe out energy band gaps and to produce localization and diffusive transport [15-18]. Previous studies on TE-polarization PBG materials were uncommon and have only focused on cases with relatively high refractive-index contrast [19-21].

In this study, we construct hyperuniform disordered photonic materials with a wall-network architecture (no cylinders), and demonstrate that TE-polarization band gaps are possible in disordered structures even with low index-of-refraction contrast (1.6:1), a regime where one might expect the disorder to disrupt band-gap formation, and where a tight-binding picture [11] no longer applies. We also show that, in a progression of structures with increasing rotational symmetry, the angular dependence of the stop bands decreases, culminating in the truly isotropic PBG of the disordered system, despite the lack of Bragg scattering.

## 2. Design of the hyperuniform disordered structure

Central to the novel designer PBG materials [13] are the concepts of hyperuniformity and stealthiness. A point pattern is hyperuniform if the number variance within a "spherical" sampling window of radius $R$, $\sigma^2(R) \equiv \langle N_R^2 \rangle - \langle N_R \rangle^2$, grows more slowly than the window volume for large $R$, i.e., more slowly than $R^d$ in $d$ dimensions [22, 23]. Note that, for many 2D random systems (Poisson distribution and molecular liquids) the number variance is proportional to the window area, $\sigma(R) \propto R^2$; whereas the number variance for hyperuniform structures, grows more slowly than the window area, e. g. for crystals, quasicrystals, and hyperuniform disordered structures used in this study $\sigma(R) \propto R$. Because of this feature, the photonic design pattern has hyperuniform long-range density fluctuations (or, equivalently, a structure factor $S(k)$ approaches zero for wavenumber $k$ approaching zero), similar to crystals [22]; at the same time, the pattern exhibits random positional order, isotropy, and a circularly symmetric diffuse structure factor $S(k)$ similar to that of a glass. The cases considered for our structures have the slowest possible rate of growth, which is proportional to $R^{d-1}$. In reciprocal space, hyperuniformity corresponds to having a structure factor $S(k)$ that tends to zero as the wavenumber $|k|$ tends to zero (omitting forward scattering), i.e., the infinite-wavelength density fluctuations vanish. In particular, for generating hyperuniform disordered point patterns, we consider "stealthy" point patterns with a structure factor $S(k)$ that is isotropic, continuous, and equal to zero for a finite range of wavenumbers $|k| < k_C$ for some positive $k_C$ [24] Stealthiness imposes density-fluctuation correlations on intermediate scales. The larger the value of $k_C$, the stealthier the point pattern is and the more intermediate-range order there is. Hyperuniform materials can then be constructed by first mapping a hyperuniform point pattern onto a network structure using a mathematical protocol [13] and fabricating the structure using materials that interact resonantly with electromagnetic, electronic, or acoustic excitations.

Our hyperuniform disordered "wall-network" structure was designed using the protocol described in [13]: after generating hyperuniform point patterns within a square of side length $L$, where $L$ is about 22 times the average inter-particle spacing $a$, we employ a centroidal tessellation of the point pattern to generate a "relaxed" dual lattice. By construction, the dual-lattice vertices are trihedrally coordinated.

We then connect the lattice vertex pairs with dielectric walls of fixed width to generate the trihedral-network photonic architecture [13].

Our simulations show that TE-polarization PBGs can exist in disordered systems that exhibit a combination of hyperuniformity, uniform local topology, and short-range geometric order, all of which occur naturally in the trihedral photonic architectures introduced above. As the point pattern becomes more hyperuniform and more stealthy, the scattering units for TE-polarized radiation (the irregularly shaped air cells) become more and more uniform in both shape and size and support better and better defined electromagnetic resonances, enabling localized electromagnetic resonances responsible for the TE PBGs. With further optimization of the wall thickness $t$ for a given dielectric contrast, the simulations predict the formation of a TE-polarization band gap, a forbidden frequency range with zero density of states.

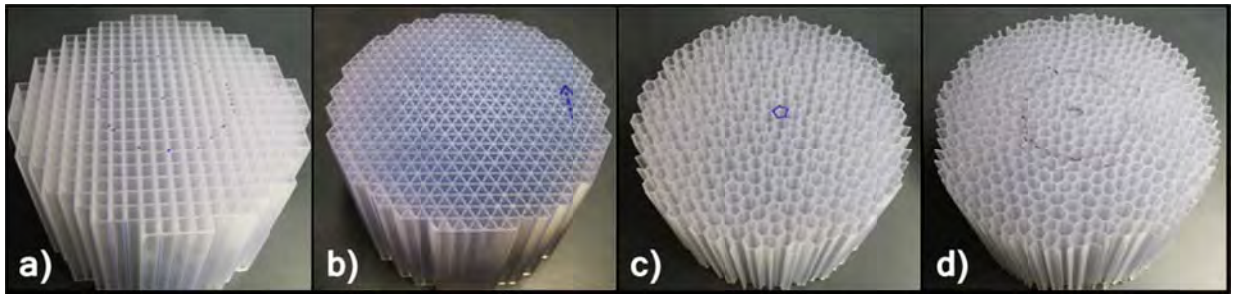

Fig.1 Photos of the samples used in this study. All structures are 100 mm high. a) The square lattice crystal (lattice spacing = 6.60 mm). b) The triangular lattice crystal with (lattice spacing = 6.60 mm). c) The five-fold symmetric quasicrystal (lattice spacing=6.58 mm). d) The hyperuniform disordered structure (average inter-vertex spacing a=5.72 mm). The volume-filling fraction is 40.5% for all the samples.

To demonstrate the angular dependence of photonic properties and the evolution with increasing rotational symmetry, we have studied the properties of four different structures. Our structures were fabricated with a stereolithography machine (model SLA-7000 from 3D Systems ®) that produces a solid plastic model by ultraviolet laser photo-polymerization. The resin used was *Accura® 60* (a clear, polycarbonate-like plastic) from 3D ® Systems Corporation. The resolution is 0.1mm in both lateral and vertical directions. Fig. 1 shows these centimeter-scale photonic structures, which are designed to have a nearly circular boundary with a diameter of 135 mm (~ 22 *a*). The five-fold-symmetry quasicrystal wall-network structure was constructed using the same centroidal tessellation protocol applied to a point pattern consisting of the vertices of a periodic approximant to a Penrose tiling, as described in [25]. The wall thickness $t$ was set for each structure separately, so that the total volume filling fraction is about 40.5% for all the samples.

3. Results and discussion

*3.1. Measurements of microwave transmission*

Measurements were made using microwave radiation in the frequency range of 15-35 GHz (0.33 to 0.77 c/a), and with a setup similar to the one described in [8]. Each sample was placed between two facing microwave horn antennas connected to a HP85210C vector network analyzer. A single polarization mode was coupled through a pair of rectangular horn antennas. The horns were placed a distance of 40*a* apart to produce approximately plane wavefronts at the samples. The samples were aligned so that the incident beam was perpendicular to the vertical axes. We rotated each sample about its vertical axis and recorded the transmission every two degrees. The transmission is defined as the ratio between detected intensities with and without the sample in place.

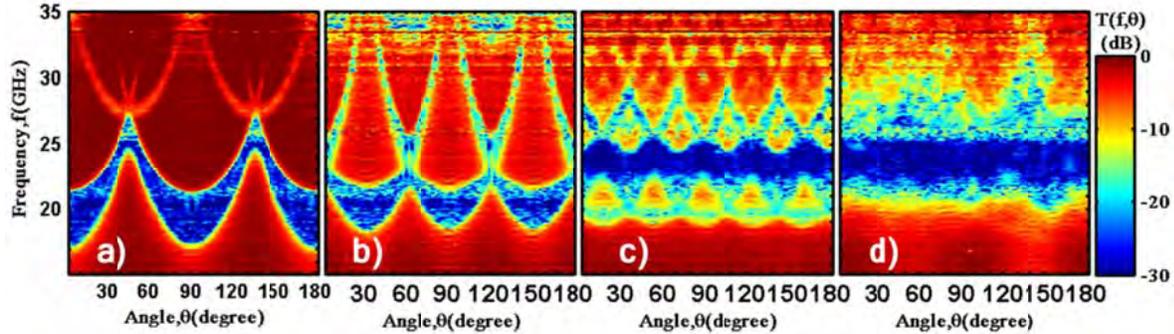

Fig. 2 Contour plots of the measured transmission as a function of frequency and incident angle. a) Square-lattice crystal. b) Triangular-lattice crystal. c) The five-fold quasicrystal. d) The hyperuniform disordered structure. For crystals or quasicrystals, stop bands (blocking regions) occur due to Bragg scattering, and their center frequency and width vary rapidly with the incident angle. In the five-fold quasicrystal, the angular difference between the most different symmetry directions is small enough to allow the blocking frequencies to overlap in all directions to form PBGs. In the hyperuniform disordered sample, a truly isotropic PBG is observed around 23.5 GHz despite the low index-contrast of 1.6:1 and the lack of Bragg scattering.

In Fig. 2, we use 3D color contour plots to present the measured transmission for all samples as a function of frequency and incident angle. The progression from left to right shows the effect of the increasing rotational symmetry of the structures. The green-to-blue regions are stop bands, and the variation of the frequency range of the stopped bands with angle prevents the formation of a PBG (blocking at all angles) for the square and triangular lattices. The more symmetric quasicrystal structure achieves narrow PBGs, and the hyperuniform disordered structure forms a truly isotropic PBG. In Fig. 3 we show the measured transmission for the four samples as a function of frequency and angle in polar coordinates, along with their calculated structure factors $S(k)$. Rotational symmetries are clearly visible in both Fig. 2 and Fig. 3. The polar-coordinate plots are especially useful for revealing the relationship between PBG formation and Bragg scattering planes (Brillouin zone boundaries) in crystals and quasicrystals.

For crystals and the quasicrystal, stop bands due to Bragg scattering change their frequency continuously as a function of the incident angle. A wavevector ***k*** that resides on the plane defined by a reciprocal lattice vector ***G*** is Bragg-scattered by ***G***. Such a wavevector hits a Brillouin-zone boundary, and its wavenumber $k=|\mathbf{k}|$ satisfies the condition $|\mathbf{k}|=|\mathbf{G}|/(2\cos(\theta))$. When the dielectric contrast is low, the center frequency of a stop band due to Bragg scattering is, to the lowest-order approximation, $f \sim c|k|/(n2\pi)$, which is inversely proportional to $\cos(\theta)$, where $c$ is the speed of light in vacuum, and $n$ is the Bruggeman effective medium index [26]. For the square lattice, the angular dependence due to its 4-fold rotational symmetry is so large that the center frequency of the measured stop bands in the X direction (0 degree) and the M direction (45 degree) vary by about 40%, making it impossible to form a bandgap (overlap of stop bands along all propagation directions) at this low dielectric contrast.

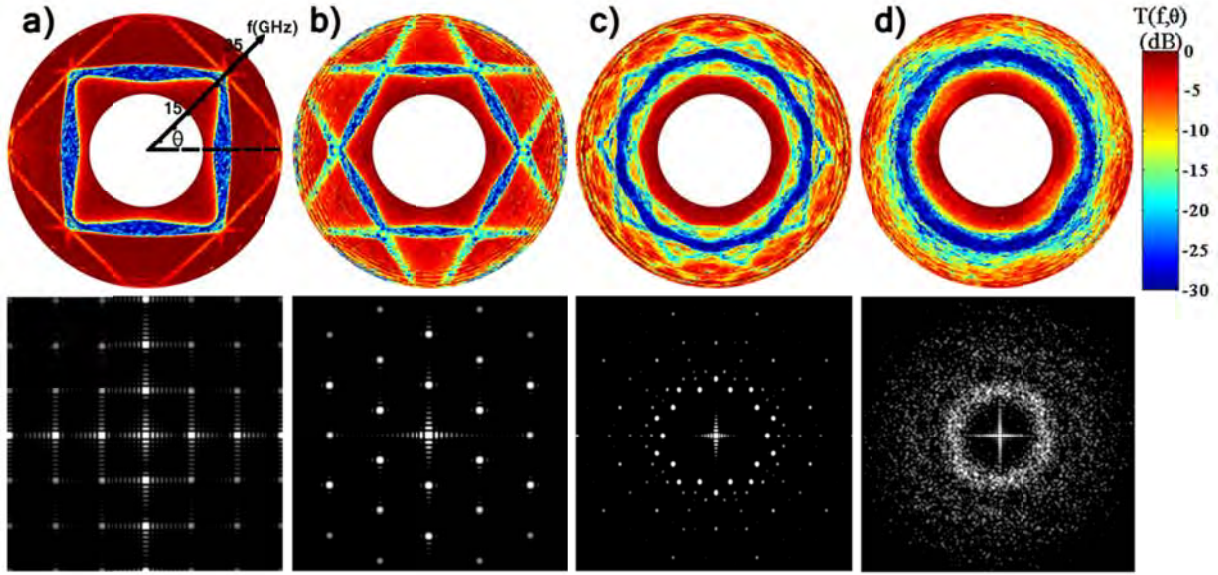

Fig. 3. Measured transmission and calculated structure factors for the four samples tested.  Contour plots (top) in polar coordinates of the measured transmission as a function of frequency, which is the distance to the pole in the radial direction (f=r, 15 to 35 GHz), and incident angle which is the angular coordinates ($\theta$, 0 to 360 degree) and calculated structure factors (bottom). a) The square crystalline lattice. b) The triangular crystalline lattice. c) The five-fold quasicrystal. d) The hyperuniform disordered structure. For the crystals and the quasicrystal, stop bands, due to Bragg scattering, are seen to occur along their Brillouin zone boundaries, directly related to the corresponding structure factors. Large angular variations prevent PBG formation at low index-contrast. The structure factor for the hyperuniform disordered structure is isotropic and continuous. It has been engineered to be equal to zero for small wavenumbers, and exhibits a broad ring of maximum values around a characteristic wavelength range. Correspondingly, a truly isotropic PBG is observed.

Moreover, in the square lattice, the stop-gap widths vary dramatically in different directions. The triangular lattice, with its 6-fold rotational symmetry, exhibits a much smaller variation in angle than the square lattice. However, the remaining angular dependence is still too large to allow the overlap of stop bands in all directions to form a bandgap. The Penrose quasicrystal has 10-fold rotational symmetry in reciprocal space, because the density is a real function and its Fourier components are equal for +$k$ and −$k$. Hence, its spectrum repeats every 36 degrees, and the change between the furthest symmetry points (from 0 to 18 degrees) is much smaller than that found in the crystalline structures, enabling PBGs to form. For the hyperuniform disordered sample, the transmission spectrum is truly isotropic without any angular dependence; hence the stop bands overlap to form a TE PBG for all directions.

In the polar-coordinate plots, the effective Brillouin-zone structures in reciprocal space can be visualized directly [8]. For crystals and quasicrystals, the fact that the low-transmission regions correspond to straight lines in Fig. 3 demonstrates that the stop bands are due to Bragg scattering and lie on reciprocal-space planes along the Brillouin zone boundaries. The calculated structure factors for the five-fold quasicrystal show two sets of 10-point bright peaks. For this decoration of the quasicrystal, the intensities of the unlimited number of fractal Bragg peaks at small $k$ values are too small to be visible at this resolution. The transmission plot of the five-fold quasicrystal sample shows two sets of ten straight lines associated with Bragg scattering planes arranged with 10-fold symmetry. Along both the two sets, it is apparent that stop bands overlap to form bandgaps. Quasicrystals are self-similar and quasiperiodic, so they can, in principle, have many band gaps separated by mean wavenumbers with irrational ratios.

As the rotational symmetry of the photonic solid increases, the corresponding stop bands exhibit less and less angular dependence and thus facilitate the formation of a PBG for all directions. For the hyperuniform disordered sample, the structure factor is engineered to be isotropic and stealthy, resulting in a truly isotropic bandgap, as seen in the rotationally invariant deep blue ring in the polar-coordinate transmission plot. This statistical isotropy, an inherent advantage of disordered structures, offers freedom for functional-defect design that is not possible for the crystal symmetries [27, 14].

*3.2. Simulations of band structure and field distributions*

In disordered structures, a mobility gap associated with localization due to disorder can also occur, but its nature is rather different from that of a true photonic band gap in which no states, propagating or localized, exist. In order to verify that, for the hyperuniform disordered structure, the transmission gap we observed is due to a real forbidden frequency region (PBG) with the absence of any photonic states, we have carried out numerical simulations to show the band structure for the four samples studied. The theoretical band-structure calculations for the five-fold quasicrystal and the disordered structures were obtained using a supercell approximation and the conventional plane-wave expansion method [28, 29]. The supercell used for the hyperuniform disordered structure is a $\sqrt{N}a \times \sqrt{N}a$ square, where $N$=500 is the number of points in the pattern and $a$ is the average point separation. The convergence of the results for larger supercell sizes has been confirmed. We solve the vectorial Maxwell equations, assuming the structure is infinitely long in the direction perpendicular to the 2D plane, and the results are presented in Figure 4. For the quasicrystals, we use the periodic approximant scheme described at length in [25]. Here, we employ the 5/3 periodic approximant, which has a rectangular shaped unit cell with $L_x$≈34.27$a$, $L_y$≈13.04$a$, containing 550 points. The high-symmetry points shown in Figure 4c and 4d are vertices of the irreducible first Brillouin zone of the supercells: $k_\Gamma$=0; $k_X$=$b_1$/2; $k_M$=($b_1$+$b_1$)2; $k_R$=$b_2$/2; where $b_1$ and $b_2$ are basis vectors of the reciprocal lattice of the supercells.

Comparing Fig. 4 with Fig. 2 clearly shows that the measured and calculated stop bands agree very well in center frequencies, width and angular dependences for the crystals, verifying the validity of our methods. In the quasicrystal and disordered cases, the number of relevant bands increases proportionally with the number of scattering elements in their supercells. For the quasicrystal sample, there is a second, lower frequency gap, whose nature is related to multiple scattering phenomena on longer length scales associated with the quasiperiodicity [30].

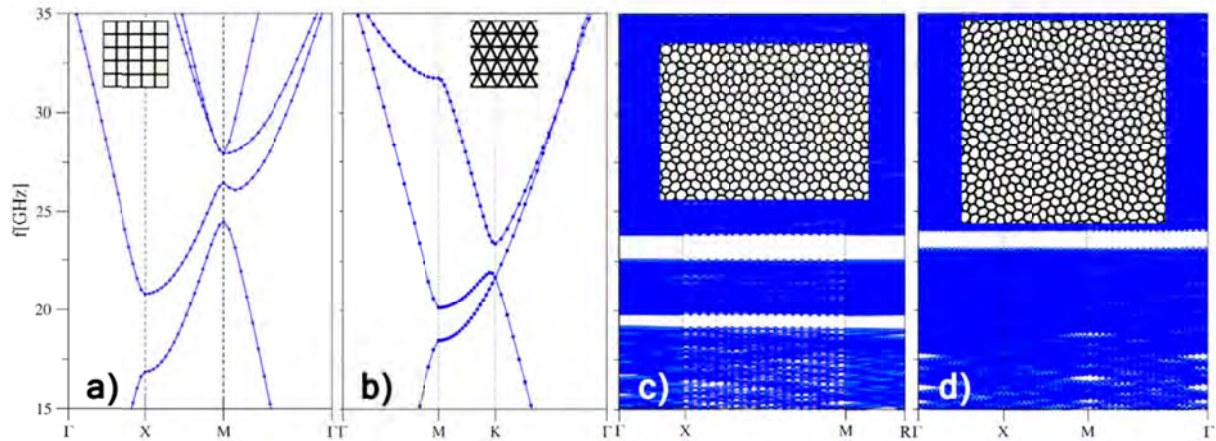

Fig. 4. Simulated band structures (blue). a) Square-lattice crystal. b) Triangular-lattice crystal. c) The five-fold quasicrystal. d) The hyperuniform disordered structure. The frequency axis covers our measurement range of 15-35 GHz. The inserts in c) and d) show a portion of the supercell used for the corresponding simulations. The calculated PBG for the hyperuniform disordered sample is from 23.1GHz to 24.1 GHz, in good agreement with the measured gap.

For the hyperuniform disordered sample, a PBG free of any photonic states also occurs, despite the lack of long-range translational order and Bragg scattering. The frequency of the calculated PBGs for the quasicrystal and hyperuniform disordered samples agrees well with the values determined from the measured transmission in Fig. 2.

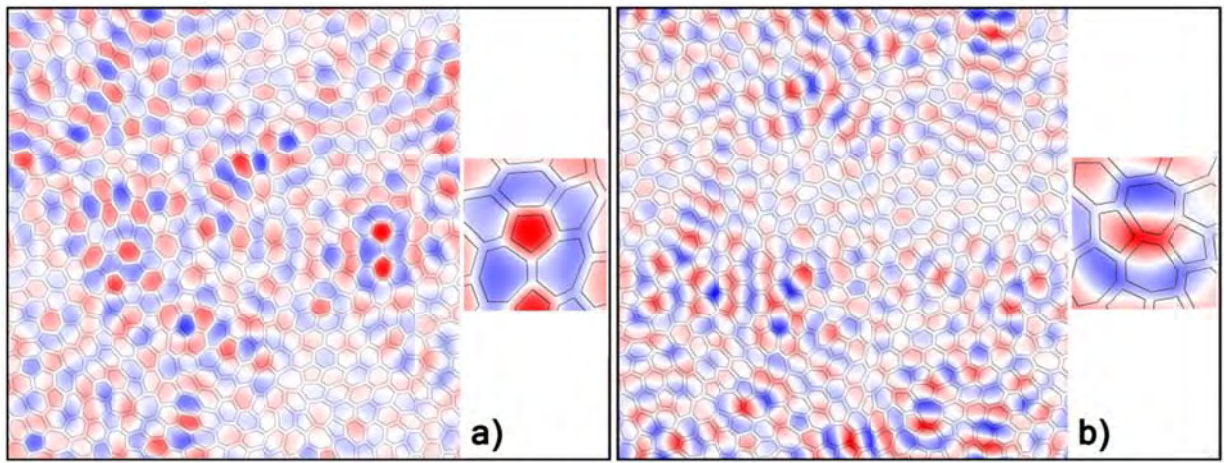

Fig 5. Simulation results of magnetic-field distribution of states near the bandgap. a) A state below the PBG. b) A state above the PBG. Zoom-in inserts show the relation of the field distribution to the cell walls. The positive (negative) values of the field are represented by the red (blue) shade and the dielectric wall boundaries are shown in black.

Further indication for the formation of a real energy bandgap in the disordered structure comes from the distribution of energy throughout the structure, above and below the bandgap. Figure 5 presents the simulation results of the azimuthal magnetic-field distribution in the hyperuniform disordered structures for two TE polarized states, below and above the bandgap. Here we show that, despite the low index-contrast, the formation of the PBG is still closely related to the formation of electromagnetic resonances localized within the network cells. For the mode below the gap (Fig. 5a), the zoom-in insert shows that the magnetic field is mostly localized inside the air cells of the network (electric fields are concentrated in the dielectric material). For the mode above the gap, the zoom-in insert shows the complementary situation, the magnetic field tending to localize in the higher-index dielectric material (electric fields are concentrated in the air cells). In between these two groups of modes (dielectric bands and air bands), there is a real energy bandgap (forbidden frequency) similar to the TE polarization PBG found in connected photonic crystal networks between the dielectric band and the air band. We next calculate the electric energy concentration factor $C_F$ defined by [28]:

$$C_F \equiv \frac{\int_{\text{high-index region}} d^2\mathbf{r}\, \varepsilon(\mathbf{r}) |E(\mathbf{r})|^2}{\int_{\text{supercell}} d^2\mathbf{r}\, \varepsilon(\mathbf{r}) |E(\mathbf{r})|^2}. \qquad (1)$$

In Fig. 6, we show the variation of the concentration factor as a function of frequency, in the region surrounding the band gap. Clearly, the large difference between the electric field distribution for modes below and above the gap, is responsible for the opening of a sizable photonic band gap in hyperuniform disordered structures. In the disordered sample, some of the modes close to the PBG are localized or diffusive. These modes contribute to the low transmission measured near the gap region.

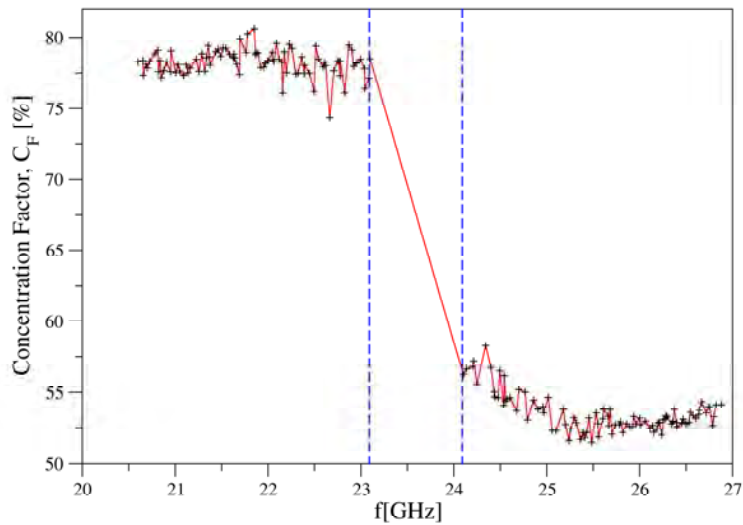

Fig 6. Calculated field concentration factor as a function of frequency. The two vertical dashed lines represent the calculated bandgap boundaries.

*3.3. Discussion*

The formation of PBGs is well understood for crystals and quasicrystals using Bloch's theorem [28]: a periodic modulation of the dielectric constant mixes degenerate waves propagating in opposite directions and leads to standing waves with high electric field intensity in the low-dielectric region for states just above the gap and in the high-dielectric region for states just below the gap. Long-range periodic order, as evidenced by Bragg peaks, is necessary for this picture to hold. Recently, the tight-binding approximation has been applied to explain apparent gaps in 3D amorphous diamond photonic structures, in analogy to amorphous Si electronic bandgaps [11]. In contrast, although our hyperuniform disordered structure does not have translational order or Bragg peaks, it still exhibits a real gap (with the absence of any photonic states) even at low index contrast, where the tight-binding approximation condition is no longer valid.

The presence of a real energy bandgap in hyperuniform disordered materials characterized by the absence of any states of a particular polarization, propagating or localized, is highly relevant for various applications, such as functional defects, i.e., cavities and waveguides [14, 27]. For the present case of symmetric 2D slabs of the desired height, polarization is preserved. Devices can be constructed for guiding or otherwise controlling either TE or TM modes. Thus the observation of an isotropic PBG for TE modes directly enables a number of applications suitable for PBG materials. The intrinsic anisotropy associated with the periodicity of photonic crystals can greatly limit the scope of PBG applications and places a major constraint on device design. For example, even though three-dimensional (3D) photonic crystals with complete PBGs have been fabricated for two decades [31] 3D waveguiding continues to be a challenge. Recently, the first successful demonstration of 3D waveguiding has been reported [32]. However, in the 3D woodpile photonic crystal reported, it was proved that waveguiding is possible only along certain major symmetry directions, due to the mismatch of the propagation modes in line defects along various orientations. The intrinsic isotropy associated with our engineered hyperuniform disordered structures offers unprecedented freedom in defect-architecture design, unlimited by crystalline or quasicrystalline symmetries, including light guiding with arbitrary bending angles along freeform paths [14, 27].

In this study, we have focused on "wall-network" structures for TE polarization PBGs. Similar design principles [13] can be applied to obtain TM-polarization PBGs [33] or complete PBGs [13, 14] and can be extended to 3D as well. These results are applicable to all wavelengths. Hence, this novel class of disordered photonic bandgap materials may be used for many applications envisioned for photonic crystals. Deep reactive ion etching on silicon can be used to construct similar hyperuniform disordered network structures with a much wider TE PBG in the infrared or optical regimes. Since this single-polarization PBG exists even at the low index contrast of 1.6:1, it becomes feasible to use soft-matter materials and self-assembly to generate similar hyperuniform disordered network structures for TE-polarization PBG, for example, by polymerizing the continuous phase of a foam or emulsion [34]. The intrinsic isotropy of these hyperuniform disordered materials offers defect design freedom, not available in photonic crystals or quasicrystals, that may play an essential role in the development of flexible optical insulator platforms and isotropic light and thermal radiation sources. It also offers advantages in PBG-enhanced technologies, e.g., displays, lasers [35], sensors [36], telecommunication devices [37], and optical micro-circuits [6].

### 4. Conclusions.

We have designed and tested an engineered hyperuniform disordered network structure with a refractive-index contrast of 1.6:1, which exhibits an isotropic TE-polarization PBG verified in both measurement and simulation. For comparison, we have studied similar "wall-network" photonic crystals or quasicrystals of square, triangular and Penrose lattices. In a series of structures with increasing rotational symmetry, culminating in our disordered structure, the photonic stop bands exhibit progressively less angular dependence, with only the disordered structure forming a truly isotopic PBG. The intrinsic isotropy of our disordered structure is an inherent advantage associated with the absence of limitations of orientational order. The combination of hyperuniformity, uniform local topology, and short-range geometric order appear to be crucial for PBG formation.

In summary, we have extended the creation of PBG media with disordered structure to low index-contrast regime, while illustrating the role of isotropy in PBG formation. This isotropy may give disordered structure significant advantages over crystalline photonic materials limited by their periodicities. Potential applications include novel architectures for cavities and waveguides displaying arbitrary bending angles [14, 27] and highly efficient thin-film solar cells [10] and light-emitting diodes [38].

### 5. Acknowledgments

This work was partially supported by the Research Corporation for Science Advancement (Grant 10626 to W. M.), the San Francisco State University internal award to W. M., the University of Surrey's support to M. F. (FRSF and Santander awards), and the National Science Foundation (NSF DMR-1105417 and NYU-MRSEC DMR-0820341 to P.M.C, DMR-0606415 to ST, and ECCS-1041083 to P.S.J and M.F.). We thank Roger Bland for helpful discussions.